
\input harvmac

\Title{HUTP-92/A062} {Will discrete symmetries help solve the
hierarchy
problem ?}
\centerline{Meng Y. Wang}
\centerline{Eric D. Carlson}
\bigskip\centerline{Lyman Laboratory of Physics}
\centerline{Harvard University}\centerline{Cambridge, MA 02138}
\medskip

\vskip.3in
We find that massless Higgs doublets at the GUT scale can be the natural
result of a discrete symmetry. Such a mechanism does not
require elaborate fine tuning or complicated particle content. The same
discrete symmetry will also protect against proton decay and flavor changing
neutral currents. However, this mechanism always predicts non-minimal
standard models. An explicit example of how this mechanism works is also
included.
\Date{10/92}

\newsec{Introduction}

Ever since the first grand unified theory was proposed 18 years
ago\ \ref\rgeorgi{H.Georgi \hbox{and} S.L.Glashow, {\it
Phys. Lett. }{\bf 8} (1964) 214.},
physicists
have been troubled by the hierarchy problem, {\it i.e.}, the existence of
two mass scales differing by approximately thirteen orders of magnitude. In
terms of perturbative field theory, this implies an almost exact cancelation
between the bare masses of the Higgs bosons and their radiative
corrections. Since no symmetry prohibits these mass terms, it seems unlikely
that such cancellations should occur. Two possible solutions have been
suggested. The first approach is to introduce new strongly coupled gauge
interaction, known as technicolor, and to make the Higgs a composite particle,
so that the hierarchy problem is automatically evaded. Unfortunately, this
type of theory is not yet well understood due to its strongly coupled nature.
We will turn our attention to the second type of theory: supersymmetry (SUSY).
Supersymmetry is a symmetry between bosons and fermions. If supersymmetry is
exact, the radiative corrections to Higgs mass terms will vanish due to the
cancelation between diagrams with boson loops and those with fermion ones.
Therefore, if its bare mass is zero at the grand unification theory (GUT)
scale, the Higgs particle will remain massless down to the SUSY breaking
scale, which can be made sufficiently small, thus solving the hierarchy
problem.

However, there remains one thorny problem in this rosy picture --- the fact
that the Higgs doublets have to be massless at the GUT scale. It may not seem
unnatural at first, since lepton superfields are also massless. However,
leptons come with color triplet partners (quarks), which are also
massless at the GUT scale, while color triplet Higgs have to have GUT masses
in order to suppress proton decay. We will refer to this problem as the
doublet-triplet hierarchy (The authors of Ref.\ \ref\rmatterp{M.C.Bento,
L.Hall and G.G.Ross, {\it Nucl. Phys. }{\bf B292} (1987) 400.} call it the
``Second Hierarchy Problem".).

Various authors have attempted to construct models that solve this problem
without fine-tuning. One approach is the so-called ``sliding singlet"
mechanism\ \ref\rsliding{E.Witten, {\it Phys. Lett. }{\bf 105B} (1981) 267
\semi L.Ibanez and G.G.Ross, {\it Phys. Lett. }{\bf 110B} (1982) 215.} in
which a singlet Higgs particle is introduced so that its vacuum expectation
value (VEV) will almost completely cancel the doublets' otherwise large
masses. Unfortunately, it was pointed out later\ \ref\rnosliding{J.Polchinski
\hbox{and} L.Susskind, {\it Phys. Rev. }{\bf D26} (1982) 3661 \semi
H.P.Niles, M.Srednicki and D.Wyler, {\it Phys. Lett. }{\bf B124} (1982) 337.}
that this solution is not really the global minimum of the effective potential
as hoped, and it is for this reason that no satisfactory models have been
successfully built based on this idea. Another ingenious method of producing
light Higgs doublets is the ``missing partner" mechanism\
\ref\rmissing{S.Dimopoulos and F.Wilczek, Santa Barbara Preprint VM-HE81-71
\semi B.Grinstein, {\it Nucl. Phys. }{\bf B206} (1982) 387 \semi
H.Georgi, {\it Phys. Lett. }{\bf B108} (1982) 283 \semi A.Masiero,
P.V.Nanopoulos, K.Tamvakis and T.Yamagida, {\it Phys. Lett. }{\bf B115} (1982)
380.}. This method employs special
group structure in the Higgs sector to give colored Higgs particles large
masses while prohibiting the doublets from getting masses by using additional
discrete symmetries. The drawback of this type of models is that they always
require very large representations, making the particle content somewhat
complicated and aesthetically unpleasant. More recently, a third method was
proposed\ \ref\rsu6{A.A.Ansel'm, {\it Sov. Phys. }{\bf JETP67-4} (1988) 663,
and references therein.} which involves a global $SU(6)$ containing the gauge
group $SU(5)$ as a subgroup. When $SU(6)$ spontaneously breaks, the doublet
Higgs become pseudo-Goldstone bosons and remain light. However, it is not
clear how to justify the introduction of this $SU(6)$ group since global
symmetries are usually considered undesirable. There have also been a number
of other ideas, but all of them seem to have fine-tuning hidden somewhere.

We will explore instead the possibility of making doublet Higgs bosons light
with a mechanism using nothing but a discrete symmetry such as the cyclic
group $C_n$. This scheme is implemented in the framework of the supersymmetric
$SU(3)^3$ GUT model\ \ref\roxford{B.Greene, K.H.Kirklin, P.J.Miron and
G.G.Ross, {\it Phys. Lett. }{\bf B180} (1986) 69 \semi {\it Nucl. Phys. }{\bf
B278} (1986) 667 \semi
{\bf B292} (1987) 606.}, also known as the trinification scheme\
\ref\reric{E.D.Carlson and M.Y.Wang, Harvard Preprint HUTP-92/A057, and
references therein.}. Such discrete symmetries are common results of string
compactification, and are needed to prohibit proton decay\ \rmatterp. However,
we will consider the general case of a SUSY $SU(3)^3$ GUT with all
phenomenologically feasible discrete symmetries and particle content, and not
restrict ourselves to the specific model proposed in Ref.\ \roxford
\ \hbox{and}\ \reric. We find that this mechanism is possible only if the
Higgs
doublet which gives down-type quarks masses differs from the one responsible
for the leptons. Therefore, this mechanism is not compatible with the minimal
supersymmetric standard model (MSSM). However, flavor changing neutral
currents (FCNC) won't be a problem because the same discrete symmetry which
generates the doublet-triplet hierarchy also prohibits FCNC.

We first describe the basic idea behind this mechanism in Section 2, and
give some examples. In Section 3, we start to construct realistic models,
and show that we require more than one pair of Higgs doublets. We then present
an example of working non-minimal models in Section 4.

\newsec{Basic ideas}

Our scheme is based on the group $SU(3)_C \times SU(3)_L \times
SU(3)_R$, where $SU(3)_C$ is the familiar color $SU(3)$, $SU(3)_L$ contains
weak $SU(2)$, and $SU(3)_R$ contains the right-handed analog of weak $SU(2)$.
This group is one of the maximal subgroups of $E_6$, with the fundamental
27---dimensional representation of $E_6$ becoming a direct sum of three
irreducible representations under trinification: $\Psi_L\ :\ (3,\bar{3},1),\
\Psi_R\ :\ (\bar{3},1,3),\ \Psi_l\ :\ (1,3,\bar{3})$, corresponding to the
quarks, the anti-quarks, and the leptons respectively. The explicit assignment
of particles is as follows\ \ref\rshelly{A.De R{\' u}jula, H.Georgi, and
S.L.Glashow, in {\it Fifth Workshop on Grand Unification}, eds. K.Kang,
H.Fried, and P.Frampton (World Scientific, Singapore,1984) 88.}:
$$\eqalign{\matrix{\Psi_L\cr (3,\bar{3},1)\cr}\
    &:\ \left(\matrix{u&d&B\cr u&d&B\cr u&d&B\cr}\right),\cr
   \matrix{\Psi_R\cr (\bar{3},1,3)\cr}\
    &:\ \left(\matrix{u^*&u^*&u^*\cr d^*&d^*&d^*\cr B^*&B^*&B^*\cr}\right),\cr
   \matrix{\Psi_l\cr (1,3,\bar{3})\cr}\
    &:\ \left(\matrix{E^o&E^-&e^-\cr E^+&{E^o}^*&\nu\cr
      e^+&\nu_R&N^o\cr}\right),\cr}$$
where $B$ is an additional superheavy down-type quark, $B^*$ is its
anti-particle, and $E$'s and $N^o$ are new superheavy leptons.

The Higgs needed to break the trinification group down to the standard model
can be put into a $(1,3,\bar{3})$ representation together with those needed to
break weak $SU(2)$. In supersymmetrized theories, they are just additional
generations of leptons. The VEV's which break $SU(3)^3$ are usually written as
$$\left(\matrix{0&0&0\cr 0&0&0\cr 0&0&v\cr}\right)\ \hbox{\rm and}\
\left(\matrix{0&0&0\cr 0&0&0\cr 0&w&0\cr}\right).$$

These VEV's are not put in by hand. Instead, they are the most general
$F$-flat direction of the superpotential after a choice of basis, assuming
only two multiplets grow VEV's (This can be arranged naturally\
\ref\rvev{M.Y.Wang and E.D.Carlson, preprint in preparation.}.).
Unfortunately, the $D$-flatness condition is
not satisfied by them. To make the symmetry breaking possible, it is necessary
to introduce the ``mirror particles", which come from the $\overline{27}$
representation of $E_6$. They transform under the trinification group like
$\bar\Psi_L\ :\
(3,\bar{3},1),\ \bar \Psi_R\ :\ (\bar{3},1,3),\ \bar\Psi_l\ :\
(1,3,\bar{3}).$ If the $(1,3,\bar{3})$ part gains the following VEV's,
$$\left(\matrix{0&0&0\cr 0&0&0\cr 0&0&v\cr}\right)\ \hbox{\rm and}\
\left(\matrix{0&0&0\cr 0&0&w\cr 0&0&0\cr}\right),$$
both $F$- and $D$-flatness will be satisfied. The number of light
generations $N_g$ will be equal to the difference between the number of
supermultiplets and that of their mirror partners, {\it i.e.}
\eqn\enumbergen{3\ =\ N_g\ =\ (\#\ of\ \Psi_x)\ -\ (\#\ of\ \bar\Psi_x),}
where $x=C,L,R$. Eq.\ \enumbergen\ also guarantees that our model is free of
anomalies.

If this had been the whole story, only the standard model fermions together
with the right-handed neutrinos and their superpartners would have remained
light, while all particles carrying the right quantum numbers for the standard
model Higgs would have become superheavy. This is just the same old
doublet-triplet hierarchy problem haunting us again.

In order to shed new light on this problem, let's put in an additional $C_N$
symmetry, ${\it H}$. For definiteness we choose $N=4$. $C_4$ has four
elements, $\{e,a,a^2,a^3\}$, where $a^4=e$, the identity. It also has 4
inequivalent irreducible representations, all one dimensional, which can be
categorized by the way they transform under the group element $a$, {\it i.e.}
$1$,$i$,$-1$, and $-i$, where $i=\sqrt{-1}$. For
example, the representation $-1$ transform under $\{e,a,a^2,a^3\}$ like
$\{1,-1,1,-1\}$, which means it is not a faithful representation.

Since the multiplets which develop the VEV's $v$ and $w$ do not necessarily
transform trivially under ${\it H}$, in general ${\it H}$ is broken together
with $SU(3)^3$ by $v$ and $w$. However, there always remains an unbroken
$C_4$, ${\it H^\prime}$, which is the product of ${\it H}$ and an
element $U_g$ of $SU(3)^3$, if we ignore the VEV's in $\bar\Phi_l$\ \rvev.
{}From the fact that $U_g$ commutes
with $SU(3)_C \times SU(2)_W \times U(1)_Y$ and that $v$ does not equal $w$ in
general, we conclude that $U_g$ must assume the following form,
$$U_g=\left(\matrix{1&&\cr &1&\cr &&1\cr}\right)_C \times
  \left(\matrix{i^\alpha&&\cr &i^\alpha&\cr &&i^{-2\alpha}\cr}\right)_L \times
  \left(\matrix{i^\beta&&\cr &i^\gamma&\cr &&i^{-\beta-\gamma}\cr}\right)_R,$$
where $\alpha$, $\beta$ and $\gamma$ are integers mod 4. Using the fact
that $v$ and $w$ must transform trivially under ${\it H^\prime}$, it is easy
to show that for any given set of $\alpha$, $\beta$ and $\gamma$, there exists
a unique choice for how the multiplets containing $v$ and $w$ should transform
under ${\it H}$. However, the reverse is not true because sets of ($\alpha$,
$\beta$, $\gamma$) related by the also unbroken $U(1)_Y$ are actually
equivalent. This means that ($\alpha$, $\beta$, $\gamma$) is equivalent to
($\alpha+1$, $\beta+4$, $\gamma-2$) since we can redefine $U_g$ by
$$U_g\rightarrow U_g\times\left(\matrix{1&&\cr &1&\cr &&1\cr}\right)_C \times
  \left(\matrix{i^1&&\cr &i^1&\cr &&i^{-2}\cr}\right)_L \times
  \left(\matrix{i^4&&\cr &i^{-2}&\cr &&i^{-2}\cr}\right)_R.$$
Hence we can always choose $U_g$ so that, say, $\alpha=0$, and we need only
consider theories with different $(\beta,\gamma)$ values.

Let's now look more closely at what happens under symmetry breaking. Denote
the number of $\Psi_l^{(x)}$ \hbox{\rm and} $\bar\Psi_l^{(x)}$ which transform
like $i^x$ under ${\it H}$ as $n_{\Psi_l}(x)$ \hbox{\rm and}
$n_{\bar\Psi_l}(x)$ respectively. After symmetry breaking, these fields will
break up into
\eqn\eproperty{\eqalign{\Psi_l^{(x)}\ \rightarrow\
          &(1,2,{1\over2},x+\alpha-\beta)+(1,2,-{1\over2},x+\alpha-\gamma)\cr
          &+(1,2,-{1\over2},x+\alpha+\beta+\gamma)+(1,1,1,x-2\alpha-\beta)\cr
          &+(1,1,0,x-2\alpha-\gamma)+(1,1,0,x-2\alpha+\beta+\gamma)\cr
          &\hbox{\rm and}\cr
         \bar\Psi_l^{(x)}\ \rightarrow\
          &(1,2,-{1\over2},x-\alpha+\beta)+(1,2,{1\over2},x-\alpha+\gamma)\cr
          &+(1,2,{1\over2},x-\alpha-\beta-\gamma)+(1,1,-1,x+2\alpha+\beta)\cr
          &+(1,1,0,x+2\alpha+\gamma)+(1,1,0,x+2\alpha-\beta-\gamma)\cr}}
respectively. Here the first number specifies the representation under color
$SU(3)$, the second number specifies the representation under weak $SU(2)$,
the third number is the hypercharge, the fourth one is the $C_N^\prime$
charge. It is clear that the number, $n_l^-(y)$, of
negatively-hypercharged $SU(2)$ doublets transforming under ${\it H^\prime}$
like $i^y$ can be expressed as
$$n_l^- (y)=n_{\Psi_l}(y-\alpha-\beta-\gamma)+n_{\Psi_l}(y-\alpha+\gamma)
+n_{\bar\Psi_l}(y+\alpha-\beta).$$
Similarly, for the positively-hypercharged doublets, we have
$$n_l^+ (y)=n_{\Psi_l}(y-\alpha+\beta)+n_{\bar\Psi_l}(y+\alpha-\gamma)
+n_{\bar\Psi_l}(y+\alpha+\beta+\gamma).$$
Now, for these doublets to gain GUT scale masses, the mass term has to look
like
$$(1,2,{1\over2},-y)\times(1,2,-{1\over2},y),$$
The mass matrix will therefore be divided into N blocks, each of size
$n_l^-(y)\times n_l^+(-y)$, where y goes from 0 to $N-1$. If
$n(y)=n_l^-(y)-n_l^+(-y)$ is negative (positive), we will have $\bigl| n(y)
\bigr|$ positively (negatively)-hypercharged light doublets transforming like
$i^{-y}$ ($i^{y}$) under ${\it H^\prime}$.
Notice that
\eqn\edifference{\eqalign{n(y)\ =\ \ \
&n_{\Psi_l}(y-\alpha-\beta-\gamma)-n_{\bar\Psi_l}(-(y-\alpha-\beta-\gamma))\cr
&+n_{\Psi_l}(y-\alpha+\gamma)-n_{\bar\Psi_l}(-(y-\alpha+\gamma))\cr
&-n_{\Psi_l}(-y-\alpha+\beta)+n_{\bar\Psi_l}(-(-y-\alpha+\beta)),\cr}}
so it is only the difference between $n_{\Psi_l}(x)$ and $n_{\bar\Psi_l}(-x)$
that counts. Define $\Delta n_l(x)=n_{\Psi_l}(x-\alpha-\beta-\gamma)-
n_{\bar\Psi_l}(-x+\alpha+\beta+\gamma)$, we can then rewrite \edifference\ \
as
\eqn\ediff{n(y)\ =\ \ \Delta n_l(y)+\Delta n_l(y+\beta+2\gamma)-
\Delta n_l(-y+2\beta+\gamma).}
There is also a constraint on the $\Delta n_l$ from \enumbergen\ , {\it i.e.}
\eqn\eng{\sum_{y=0}^{N-1}\Delta n_l(y) = 3.}

With the singlets, $e^+$ and $\bar e^-$, the story is much simpler. Because
the GUT scale masses can come only from terms in the superpotential that look
like $\Psi\Psi\Psi$ and $\Psi\bar\Psi$, and the first term cannot
contribute in the case of singlets, there will be just as many light
singlets as $\Delta n_l(y)$; they transform like $i^{y-3\alpha-2\beta-
\gamma}$. Since no negatively charged singlets have been observed, we have an
additional constraint,
\eqn\econstraint{\Delta n_l(y) \ge 0.}

With \ \ediff\ ,\ \eng\ and\ \econstraint\ \ in mind, we find that if, for
example, $(\alpha, \beta, \gamma)=(0,1,1)$ and $\Delta n_l(0,1,2,3) =
1,0,2,0$\ , we will have three generations of light lepton doublets, $l^-$,
transforming under ${\it H^\prime}$ like $i^0$, $i^2$ \hbox{and}
$i^2$, three
generations of light anti-lepton singlets, $e^+$, transforming like $i^1$,
$i^3$ \hbox{and} $i^3$, together with two light Higgs doublets, $h_1^+$
\hbox{and}
$h_2^-$, both transforming like $i^3$. Notice that terms such as $\ e^+ l^-
h_2\ $ necessary for lepton masses at the weak scale are present. Thus we have
successfully reproduced the lepton and Higgs sector of the MSSM.

\newsec{Models with one pair of Higgs doublets}

Now we can generalize to $C_N$ symmetry. It is actually very
straightforward. We simply replace $i$ with $i_N$, the $N$'th root of $1$, and
all the statements in the previous section remain true. It turns out that for
any $N \ge 4$, we can always find a number of models, {\it i.e.} sets of
$\alpha$,
$\beta$, $\gamma$ and $\Delta n_l(y)$'s, which give the correct lepton and
Higgs
content of the MSSM. However, when we also take into account the
quark sector, no successful model is possible. To see why this is true, let's
investigate the constraints coming from the quarks.

The analysis is parallel to that of the leptons in the previous
section. We first write down the quark analog of \eproperty,
\eqn\eproquark{\eqalign{\Psi_L^{(x)}\ \rightarrow
   &\ (3,2,{1\over6},x-\alpha)_{(u,d)}+(3,1,-{1\over3},x+2\alpha)_B,\cr
  \bar\Psi_L^{(x)}\ \rightarrow
   &\ (\bar 3,2,-{1\over6},x+\alpha)_{(\bar u,\bar d)}
    +(\bar 3,1,{1\over3},x-2\alpha)_{\bar B},\cr
  \bar\Psi_R^{(x)}\ \rightarrow
   &\ (3,1,{2\over3},x-\beta)_{\bar u^*}+(3,1,-{1\over3},x-\gamma)_{\bar d^*}
    +(3,1,-{1\over3},x+\beta+\gamma)_{\bar B^*},\cr
  \Psi_R^{(x)}\ \rightarrow
   &\ (\bar 3,1,-{2\over3},x+\beta)_{u^*}
    +(\bar 3,1,{1\over3},x+\gamma)_{d^*}
    +(\bar 3,1,{1\over3},x-\beta-\gamma)_{B^*}.\cr}}
Again mass terms can only come from $\Psi\Psi\Psi$ and
$\Psi\bar\Psi$. The first does not contribute to ($u$,$\bar u$),
($u^*$,$\bar u^*$) and ($d$,$\bar d$), so they are analogous to the
($e^+$,$\bar e^-$) pair. Meanwhile ($B$, $d^*$, $B^*$) and their mirror
particles are analogous to the lepton doublets because $\Psi\Psi\Psi$ does mix
them together.
Defining $\Delta n_{L(R)}(y)=n_{\Psi_{L(R)}}(y)-n_{\bar \Psi_{L(R)}}(-y)$,
we find again only $\Delta n_{L(R)}(y)$ matters. The two constraints, \eng\
and \econstraint, still hold except that they now read
\eqn\econquark{\eqalign{& \sum_{y=0}^{N-1}\Delta n_{L(R)}(y) = 3,\cr
 & \Delta n_{L(R)}(y) \ge 0.\cr}}

Eq.\ \econquark\ implies that the solutions for $\Delta n_{L(R)}(y)$'s fall
into three categories,\ $\left\{3,0,0,0,\dots\right\}$\ ,\
$\left\{2,1,0,0,\dots\right\}$\ and\ $\left\{1,1,1,0,\dots\right\}$\ . In the
two latter cases, the quark mass matrices are reducible, introducing unwanted
chiral $U(1)$ symmetries which remain unbroken at the weak scale. From this
analysis we can conclude

$$\Delta n_{L(R)}(y) = \cases{3, &for $y=y_{L(R)}$,\cr 0,&otherwise,\cr}$$
where $y_{L(R)}$ is a certain integer between 0 and $N-1$. We see immediately
that there are three light up-type anti-quarks transforming like
$i_N^{y_R+\beta}$, and three light quark doublets transforming like
$i_N^{y_L-\alpha}$.

We know from proton lifetime that there can be only three more light color
triplets other than what we have just named above, which will be identified as
the anti-particles of down-type quarks. This means that the three extra $B$
quarks left over from the $B\bar B$ coupling have to couple with either $d^*$
or $B^*$ to gain GUT scale masses. This in turn implies that either\
$-y_L-2\alpha = y_R+\gamma$\ or $-y_L-2\alpha = y_R-\beta-\gamma$, {\it i.e.}
\eqn\ea{y_L+y_R=-2\alpha-\gamma,}
\centerline{or}
\eqn\eat{y_L+y_R=-2\alpha+\beta+\gamma.}

Let's first consider the case of\ \ea.

The Higgs doublet $h_2^-$ which gives mass to down quarks has to have coupling
$dh_2^-B^*$, so it must transform like
$i_N^{-(y_L-\alpha)-(y_R-\beta-\gamma)}$,
{\it i.e.} $i_N^{3\alpha+\beta+2\gamma}$. Similar arguments give the ${\it
H^\prime}$ property of $h_1^+$ as $i_N^{3\alpha-\beta+\gamma}$, so
$$n(-3\alpha+\beta-\gamma)=-1.$$
Together with \ediff\ this gives
\eqn\eaa{\Delta n_l(3\alpha+\beta+2\gamma) \ge 1+\Delta
n_l(-3\alpha+\beta-\gamma).}
But there are $\Delta n_l(3\alpha+\beta+2\gamma)$\ \ $e^+$'s transforming like
$i_N^{-\beta+\gamma}$. {\it Now if $h_2^-$ also gives mass to leptons}, we
will also need $\Delta n_l(3\alpha+\beta+2\gamma)$\ \ $l^-$'s transforming
like
$i_N^{-3\alpha-3\gamma}$. Again from \ediff\ , we have
\eqn\eab{\eqalign{n(-3\alpha-3\gamma)&=\Delta n_l(-3\alpha-3\gamma)+\Delta
    n_l(-3\alpha+\beta-\gamma)-\Delta n_l(3\alpha+2\beta+4\gamma)\cr
   &\ge\Delta n_l(3\alpha+\beta+2\gamma).\cr}}
Using \eaa\ we see that
$$\Delta n_l(-3\alpha-3\gamma) \ge 1.$$
But again we must have just as many $e^+$'s transforming like
$i_N^{-6\alpha-2\beta-4\gamma}$. Once more, this implies we should have
$\Delta n_l(-3\alpha-3\gamma)$\ \ $l^-$'s transforming like
$i_N^{3\alpha+\beta+2\gamma}$, but remember that we also have an $h_2^-$
transforming exactly the same way, so therefore :
$$n(3\alpha+\beta+2\gamma)\ \ge \Delta n_l(-3\alpha-3\gamma)+1.$$
Using \ediff, this can be rewritten as
\eqn\eac{\eqalign{&\Delta n_l(3\alpha+2\beta+4\gamma)
     +\Delta n_l(3\alpha+\beta+2\gamma)-\Delta n_l(-3\alpha+\beta-\gamma)\cr
    &\ge \Delta n_l(-3\alpha-3\gamma)+1.\cr}}
There is obviously no solution which satisfies both \eab\ and \eac.

The same conclusion can be drawn for the case of \eat\ in much the same
way; hence we have demonstrated that our mechanism is not compatible with
the MSSM if the discrete group is $C_N$.

If we generalize the group to $C_{N_1} \times C_{N_2} \times \dots$,
everything we have said in this section remains valid. The only
difference is that the variables $x$,\ $y$,\ $\alpha$,\ $\beta$,\ $\gamma$\
are
now vector-like with the first component corresponding to $C_{N_1}$, etc.

The generalization to non-abelian discrete group is more interesting, although
it won't work either. Notice that an irreducible CKM matrix actually demands
that the light quarks and anti-quarks be in one-dimensional representations of
the discrete group. The Higgs also have to be in one-dimensional
representations because we want minimal models, and then the same proof goes
through.

So far we have considered exclusively so-called ``phase symmetry", {\it i.e.}
the
fields gain phases when operated on by the group. Another type of
representation is so-called ``permutation symmetry", where the fields are
transformed into each other. Fortunately all symmetries of this kind which are
not equivalent to phase symmetries will be broken by the VEV's $v$ and $w$,
thus becoming irrelevant. Therefore our mechanism is not compatible with
the MSSM for {\it any} discrete group.

\newsec{Realistic models with more Higgs doublets}

We are now ready to look at realistic models. We already know that they must
contain at least two pairs of Higgs. That translates into four Higgs
doublets, each with its own ${\it H^\prime}$ property so that we won't run
into trouble with FCNC. Together with three lepton doublets and three
singlets, we can estimate roughly that about ten representations are needed. A
simple computer routine finds two dozen such models with 3 pairs of Higgs
using the group $C_{12}$. Readers should be warned that it has not been our
intention to provide an exhaustive list of such models.

We will give just one of these models here as an example. With $N=12$, we
choose ($\alpha$, $\beta$, $\gamma$) to be (0,2,1), which is of course
equivalent to eleven other sets of values. $\Delta n_l(x)$ is set to $1$
for $x=3$, $8$, $10$; it is $0$ otherwise. This leads to three light lepton
doublets transforming like $i_N^4$, $i_N^6$ and $i_N^{11}$; three
singlets of $i_N^3$, $i_N^5$ and $i_N^{10}$; three $h_1^+$ of
$i_N^3$,
$i_N^7$ (gives up-quark mass) and $i_N^{10}$; and three $h_2^-$ of
$i_N^3$ (gives lepton mass), $i_N^8$ (gives down-quark mass) and
$i_N^{10}$. Notice that the potentially troublesome coupling
$e^+h_2^-h_2^-$
is nicely forbidden by ${\it H^\prime}$.

The quark sector is the same as we have discussed in the paragraph following
\econquark, with a constraint on the choice of $y_{L(R)}$,
$$y_L+y_R=4,$$
where $y_L$ can assume any value.

It is obvious that only one pair of Higgs couple to the quarks. Thus flavor
changing neutral currents will not be a problem.

Notice that this mechanism works for most discrete groups with more than ten
representations of dimension one. For example, it will work for the group
$C_2\times C_2\times C_3$ which appears naturally in Ref.\ \roxford\ ,
although
the particular model in the said reference does not exhibit this mechanism.

\newsec{Remarks}

We have shown that the triplet-doublet hierarchy problem can be solved in a
natural way with a discrete symmetry. We have also established that at least
two pairs of Higgs are required. If nature should indeed choose this
mechanism, it would be very difficult to further determine the details of the
mechanism such as the group $C_N$, $\Delta n$, etc., without a complete
knowledge of SUSY breaking. In other words, at this stage all the ``good"
models seem to be equally successful in most respects. One exception is the
number of Higgs because it will affect how coupling constants run. It is well
known that the three coupling constants intersect each other at one point in
the MSSM\ \ref\rrunning{W.de Boer and H.F{\" u}rstenau, {\it Phys. Lett. }{\bf
B260} (1991) 447.}. Additional Higgs particles will surely spoil this
nice result. Fortunately $SU(3)^3$ is not a real GUT group in the rigorous
sense. It still allows three distinct coupling constants even if one goes
above the symmetry breaking scale $v$. These coupling constants will be united
instead at the supposed string scale $s$. With two pairs of Higgs, it turns
out that $s$ can be easily made small enough so as to coincide with the
generally accepted value. With three pairs, $s$ tends to be higher than
desired.

\bigbreak\bigskip\bigskip\centerline{{\bf Acknowledgements}}\nobreak
We thank S.Coleman, H. Georgi and B.Greene for discussions.
M.W. would like to thank S.S.Sethi for his suggestions. This
research is supported in part by the National Science Foundation under grant
\#PHY--87--14654, and the Texas National Research Laboratory Commission under
Grant \#RGY9206.

\vfill\eject
\listrefs
\bye